\documentclass[10pt,twocolumn,british,tightenlines,eqsecnum,floats,aps,amsmath,amssymb,nofootinbib,superscriptaddress,prd,showpacs]{revtex4}
\usepackage[utf8]{inputenc}
\usepackage{color}
\usepackage{babel}
\usepackage{amsmath}
\usepackage{amssymb}
\usepackage{graphicx}
\usepackage{esint}
\usepackage[unicode=true,pdfusetitle,
 bookmarks=true,bookmarksnumbered=false,bookmarksopen=false,
 breaklinks=false,pdfborder={0 0 1},backref=false,colorlinks=false]
 {hyperref}

\makeatletter
\@ifundefined{textcolor}{}
{%
 \definecolor{BLACK}{gray}{0}
 \definecolor{WHITE}{gray}{1}
 \definecolor{RED}{rgb}{1,0,0}
 \definecolor{GREEN}{rgb}{0,1,0}
 \definecolor{BLUE}{rgb}{0,0,1}
 \definecolor{CYAN}{cmyk}{1,0,0,0}
 \definecolor{MAGENTA}{cmyk}{0,1,0,0}
 \definecolor{YELLOW}{cmyk}{0,0,1,0}
}

\usepackage{babel}
\usepackage{babel}

\makeatother

\begin{document}

\title{Semiclassical dynamics of horizons in spherically symmetric collapse}

\author{Yaser Tavakoli}
\email{tavakoli@ubi.pt}
\affiliation{Departamento de F\'{i}sica e Centro de Matem\'atica e Aplica\c{c}\~oes (CMA-UBI), Universidade da Beira Interior, 6200 Covilh\~a, Portugal}

\author{Jo\~ao Marto}
\email{jmarto@ubi.pt}
\affiliation{Departamento de F\'{i}sica e Centro de Matem\'atica e Aplica\c{c}\~oes (CMA-UBI), Universidade da Beira Interior, 6200 Covilh\~a, Portugal}

\author{Andrea Dapor}
\email{andrea.dapor@fuw.edu.pl}
\affiliation{Instytut Fizyki Teoretycznej, Uniwersytet Warszawski, ul. Ho\.{z}a 69, 00-681 Warsaw, Poland}

\begin{abstract}

In this work, we consider a semiclassical description of the spherically symmetric
gravitational collapse with a massless scalar field. In particular,
we employ an effective scenario provided by \emph{holonomy} corrections
from loop quantum gravity, to the homogeneous interior spacetime.
The singularity that would arise at the final stage of the corresponding classical
collapse, is resolved in this context and is replaced by a bounce.
Our main purpose is to investigate the evolution of trapped surfaces
during this semiclassical collapse. Within this setting, we obtain a threshold radius
for the collapsing shells in order to
have horizons formation. In addition, we study the final state of the
collapse by employing a suitable matching at the boundary shell
from which quantum gravity effects are carried to the exterior
geometry.

\end{abstract}


\maketitle

\section{Introduction}

There are two main aspects related to the final state of gravitational
collapse of a star. The first one is the singularity formation; it
is understood in the sense that as the radius of a star vanishes,
the matter energy density diverges at its center. The second is the
evolution of horizons during the collapse. In the latter, if trapped
surfaces form as the collapse proceeds, then, the final singularity
will be covered by a horizon and hence, a black hole can form. Otherwise,
if such trapped surfaces do not form as the collapse evolves, the
radial null geodesics emerging from the singularity can reach the
distant observer and the singularity will be naked \cite{RPenrose,RPenrose2,RPenrose2b,Hawking,2}.

It is believed that the singularity problem will be overcome in a
quantum theory of gravity. Loop quantum gravity (LQG) \cite{LQG,LQG2,LQG3}
is a non-perturbative and background independent approach of quantum
gravity that provides a fruitful ground to investigate the removal
of singularities \cite{Bojowald-Book}. Nevertheless, LQG, in its own form is extremely complex and difficult to directly apply. Most solutions and results in general relativity are obtained with approximations
or assumptions, one of the most widely used being symmetry reduction;
this allows to access the most interesting gravitational phenomena such as spherically symmetric gravitational collapse.
Similarly, the symmetry reduction is expected to simplify many problems of
the full quantum gravity, which provides a simple arena to test ideas and constructions introduced in the full LQG.
Moreover, by systematic perturbation expansions around
symmetric models, the crucial physical issues facing LQG can be analyzed without
restricting the number of degrees of freedom.
Loop quantum cosmology (LQC), 
being a symmetry reduced model of LQG, inherits the quantum schemes originated from
LQG that dealt with the isotropic and homogeneous universe firstly and then extended to
the inhomogeneous and anisotropic model \cite{Bojowald:2002}.
It also presents itself as a possible path to unveil the cosmological and astrophysical riddles.
Results from LQC leads to the
conclusion that the cosmological singularities are resolved in quantum
gravity \cite{A-B-L}. However, in view of the full theory, there is still considerable ambiguity and none of those results are fully satisfactory  \cite{Thiemann2006,Thiemann2006b}.

Within the context of LQC, the status of the classical
singularities that arise at the late time stages of the spherically
symmetric gravitational collapse, has been studied in LQG \cite{4,homo-Boj,TachL,RTibrewala,RTibrewala-b,Modesto-1,Modesto-1b}.
Therein, different fields, such as the standard scalar \cite{4,homo-Boj,VHussain}
or the tachyon \cite{TachL,TachClass}, have been considered to play
the role of the collapsing matter source. By employing
the quantum gravity effects (such as the inverse triad correction
imported from LQC), it was shown that the geometry of spacetime near
the classical singularity is regular. Furthermore, some novel features
such as evaporation of horizons in the presence of quantum gravity
effects were studied in Refs.~\cite{4,VHussain}. In addition, it
was shown in the Ref.~\cite{homo-Boj} that (inverse triad modifications)
quantum gravity effects predict a critical threshold scale for horizons
formation which may lead to the formation of very small nonsingular
astrophysical black holes.

In recent years, some studies of the improved LQC dynamics framework
with a massless scalar field have been developed \cite{A-P-S,A-P-S2006}
(see also Ref.~\cite{LQC-AS}). Concerning the physical implications
of the singularity resolution in LQC, it was shown that the classical
big bang singularity can be resolved and replaced by a quantum bounce
\cite{A-P-S2006,A-P-S}. In view of these elements, it is expected
that the singularity arising at the end state of gravitational collapse
could also be resolved and replaced by a bounce. However, the question
we address is how loop quantum effects can indeed affect the emergence
of trapped surfaces in this kind of models.

A trapped surface, in classical general relativity, is defined as
a compact 2-dimensional smooth space-like submanifold of the spacetime
such that the families of outgoing, as well as ingoing, future pointing
null normal geodesics are contracting \cite{RPenrose}. When the quantum
geometry regime becomes relevant, the geodesic description becomes
inaccurate and the classical statements, based on the properties of
geodesics in differential geometry, are not valid anymore and should
be replaced by something more appropriate in the context of a quantum
theory. However, in order to extract some limited physical information
out of a full quantum theory it can be of interest to employ first
the effective theory of LQG. The quantum dynamics of LQC, including
the holonomy corrections, can be approximated by a set of effective
continuous equations of motion, which results in an effective theory
of LQC \cite{A-P-S}. The effective theory takes the form of a classical
theory supplemented with correction terms inherited from the quantum
theory. Furthermore, in the semiclassical regime, the use of similar
concepts (such as singularities and trapped surfaces) of general relativity,
make the question on how the (holonomy) corrections affect this concepts
an interesting topic.

In order to achieve that purpose, we apply the recent
results of effective theories of LQC to the resolution
of singularities arising in the gravitational collapse
of stars. More precisely, herein this paper we consider a spherically
symmetric framework for the gravitational collapse whose matter content
includes a scalar field. We study a semiclassical scenario of LQC
for our collapsing model which is provided by the holonomy corrections;
in this scenario the classical singularity is resolved and is replaced
by a bounce. Then, by considering the physical conditions for trapped
surface formation, we investigate how this semiclassical modification
affects the collapse end state. Our main concern is whether the expected
final bounce can be observed by a distant observer.

The content of this paper is organized as follows. In section \ref{class},
we provide the background scenario through the choice
of a suitable spacetime geometry for the collapsing system and its
matter content. In particular, we consider a flat Friedmann-Lemaitre-Robertson-Walker
(FLRW) interior spacetime to be matched to a generalised Vaidya geometry
at the boundary of matter. The matter source is considered to be a
homogeneous and massless scalar field. In section \ref{LQG-Collapse},
we study the semiclassical scenario for the interior spacetime by
employing the holonomy corrections imported from LQC. In this section,
we also investigate how quantum effects influence the evolution of
trapped surfaces as the collapse evolves. In section \ref{Collapse-outcome},
by employing the matching conditions at the boundary of collapsing
cloud, we study the exterior geometry; depending on the initial conditions
of the collapse, we have scenarios where the final bounce is visible
to a distant observer or is covered by a nonsingular black hole horizon.
Finally, we present the conclusion of our results in section \ref{discussion}.

\section{Gravitational collapse with a scalar field}
\label{class}

We consider a spherically symmetric model for the gravitational collapse to investigate the LQC effects on the removal of the black hole singularity forming at the collapse end state. 
Techniques to handle inhomogeneous systems \cite{RTibrewala,RTibrewala-b} are still under development \cite{inhomogeneous1,inhomogeneous2,inhomogeneous3} (see also Ref.~\cite{Bojowald:2011}), but they do not easily reveal the physical picture. 
We, therefore, follow the literature (see, e.g., Refs.~\cite{4,homo-Boj,TachL}) and consider a simple toy model by taking a \emph{homogeneous}  interior spacetime for a collapsing spherical
body (star)  filled with a massless scalar field. 
In order to describe the whole spacetime
structure,  our  \emph{interior}  region 
must be matched to 
a suitable (inhomogeneous) exterior geometry at the boundary surface with the radius coordinate $r=r_{b}$.
Classically, this model always produces a black hole, but we show that holonomy corrections from LQC change this situation dramatically \cite{2}.

The matter is confined in a spherically symmetric region whose coordinates are considered to be $(t,r,\theta,\phi)$. 
The geometry in the interior region can be, in general, described by the metric \cite{LL1}
\begin{equation}
ds^{2}\ =\ -e^{2\nu(t,r)}dt^{2}+e^{2\psi(r,t)}dr^{2}+a^{2}(t)r^2 d\Omega^{2},\label{SSG-metric}
\end{equation}
where $a(t)$ is the scale factor and $d\Omega^{2}$ is the standard line element on the unit two sphere. We can identify any shell with the corresponding coordinate
radius $r$.
Considering the class of scalar fields having one timelike and three
spacelike eingenvectors, this metric can be simplified and a particular class of solutions is the
the flat FLRW \cite{SS-1,SS-1b,SS-1c}:
\begin{equation}
g_{\mu\nu}^{-}dx^{\mu}dx^{\nu}\ =\ -N^2(t)dt^{2}+a^{2}(t)\left(dr^{2}+r^{2}d\Omega^{2}\right),
\label{INTmetric}
\end{equation}
where $N(t)$ is the lapse function.
In  the case where the scalar field gradient remains always timelike (behaving like a stiff perfect fluid) and in particular when $\phi(t)\neq 0$, the scalar field collapse is continual and the endstate is singular \cite{Giambo}. 
Notice that if the scalar field $\phi$ is monotonic in $t$, it can be used as a physically meaningful parameter (e.g. the \emph{physical time}) for the collapse evolution. 
%Following the literature, we call it the \emph{physical time}. 
The interior metric (\ref{INTmetric}) was also shown to be adequate to obtain an exact soluble solution for loop quantum cosmology (sLQC) \cite{sLQC}. 
Therefore, in this toy model, we aim to follow verify the main features of the LQC approach in a gravitational collapse context. 
%%%%%%%%%%%%%%%%%%%%%%%%%%%%%%%%%%%%%%%%%%%%%%%%%%%%%
 The physical radius of such a shell is given by 
\begin{equation}
R(t,r):=a(t)r,\label{AreaRadius}
\end{equation}
known as the \emph{area radius}. In the context of the canonical analysis,
which differentiates the roles of $r$ and $t$, it is reasonable
to fix the coordinate $r$ and regard $R(t,r)$ as a function defined
in the gravitational phase space.

In order to  discuss, in the next section, the LQG  corrections
to the classical evolution, we need to replace the phase space
variables of the collapsing spacetime with Ashtekar-Barbero variables \cite{LQG,LQG2,LQG3} $(A^{i}_{a},\:E^{a}_{i})$.
After the symmetry reduction of the interior FLRW type (\ref{INTmetric}), the full phase space of gravity, $\Gamma_{\rm grav}$, is further reduced to $\Gamma_{\rm grav}^{s}$. The replacement of the phase space variables carry another level of reduction consisting in imposing a gauge fixing to the diffeomorphism freedom that essentially  reduces SU(2) variables $(A^{i}_{a},\:E^{a}_{i})$ to U(1) ones.
The reduced phase space $\Gamma_{\rm grav}^{s}$ is two dimensional and coordinatised by the new
variables \cite{A-B-L} $c:=\gamma\dot{a}$ and ${\sf p}:=a^{2}$ which are, respectively,
the conjugate connection and the triad satisfying the nonvanishing
Poisson bracket $\{c,{\sf p}\}=8\pi G\gamma/3$; moreover, $G$ is
the Newton constant, $\gamma\approx0.23$ is the Barbero-Immirzi dimensionless
parameter, and a `dot' denotes the differentiation with respect to
the proper time $t$. Notice that, for any collapsing shell labeled by $r$,  the area radius  is a (gravitational) phase space function; $R=r\sqrt{|{\sf p}|}$.

Therefore, the corresponding classical Hamiltonian constraint, obtained after this process of simplification, for the interior
geometry is provided by\footnote{Notice that, variation with respect to $N$ forces the Hamiltonian constraint (\ref{Hamil-gr}) to be zero; vanishing Hamiltonian is equivalent to the Friedmann equation,
thus, the lapse function $N(t)$ does not play a dynamical role and correspondingly does not appear in
the Friedmann equation.} \cite{A-B-L} 
\begin{equation}
C\ =\ -\frac{3}{4\pi G\gamma^{2}}c^{2}\sqrt{|\mathsf{p}|}+C_{{\rm matt}}\ .\label{Hamil-gr}
\end{equation} 
(Notice that for a collapsing model, since $\dot{a}<0$
hence, $c<0$). The interior matter content is assumed to be  a massless scalar field whose Hamiltonian reads 
\begin{equation}
C_{{\rm matt}}=\rho V\ =\ \frac{\pi_{\phi}^{2}}{|{\sf p}|^{3/2}}\ ,
\end{equation}
where $V=|{\sf p}|^{3/2}$ is the volume of the fiducial cell \cite{A-B-L}.
For a massless scalar field $\phi$, the energy density $\rho$ and
the pressure $p$ coincide, and can be expressed in terms of the matter
dynamical variables as $\rho=p=\pi_{\phi}^{2}/2|{\sf p}|^{3}$. Notice
that, since the expression of the constraint equation (\ref{Hamil-gr})
does not depend on the the scalar field $\phi$, its momentum $\pi_{\phi}$
is a constant of motion; moreover, the matter field $\phi$ and its
conjugate momentum $\pi_{\phi}$ satisfy the Poisson bracket $\{\phi,\pi_{\phi}\}=1$,
thus, they can be used as the coordinates of a two dimensional matter
phase space.

By solving for the Hamiltonian constraint (\ref{Hamil-gr}), the corresponding
Einstein's equations for the interior region can be presented \cite{2}
as 
\begin{equation}
8\pi G\rho=\frac{F_{,r}}{R^{2}R_{,r}}\ ,\ \   \  \  \ 8\pi Gp=-\frac{\dot{F}}{R^{2}\dot{R}}\ ,\ \  \  \  \ \dot{R}^{2}=\frac{F}{R}\ ,\label{einstein-1}
\end{equation}
where the `$,r$' denotes the differentiation with respect to the
coordinate $r$. The \emph{mass function} $F(t,r)$ is defined to
be the total gravitational mass within the shell labelled by $r$. By integrating the first relation in Eq.~(\ref{einstein-1}) we can write the mass function as
%\textcolor{red}{[write $F$ as a function of phase space variables; $F=(8\pi G/3)\rho R^3$]}
\begin{equation}
F(r,t)\ =\ \frac{8\pi G}{3}\rho R^{3}\  .
\label{mass-class}
\end{equation}
We can also write the mass function (from the last relation in Eq. (\ref{einstein-1})) as a function of the phase space variables $(c, {\sf p}, \phi, \pi_\phi)$:
\begin{equation}
F(c, {\sf p})\ =\   \frac{r^3}{\gamma^2} c^2\sqrt{|{\sf p}|}\ = \  \frac{8\pi Gr^3}{3} \frac{\pi_{\phi}^{2}}{|{\sf p}|^{3/2}} \  .
\label{mass-class}
\end{equation}

In order to investigate the geometry of trapped surfaces inside the
star, it is convenient to study the behaviour of the radial null geodesics
emerging from the interior spacetime. Introducing the null coordinates  \cite{MS} 
\begin{align}
d\xi^{+} & \ =\  -\frac{1}{\sqrt{2}}\left[Ndt-a(t)dr\right], \notag \\ 
d\xi^{-}& \ =\ -\frac{1}{\sqrt{2}}\left[Ndt+a(t)dr\right],\label{doublenull}
\end{align}
the interior metric (\ref{INTmetric})
can be transformed into the double null form \cite{MS} 
\begin{equation}
g_{\mu\nu}^{-}dx^{\mu}dx^{\nu}=-2d\xi^{+}d\xi^{-}+R^{2}d\Omega^{2}.\label{metricdnull}
\end{equation}
Consequently, the radial null geodesics are obtained through the solution of $g_{\mu\nu}^{-}dx^{\mu}dx^{\nu}=0$,
and assuming the condition that $d\Omega^{2}=0$. Subsequently, we
deduce that there exists two kinds of null geodesics corresponding
to $\xi^{+}=const.$, and $\xi^{-}=const$. At this point we are able
to compute the \emph{expansion parameters} \cite{MS} 
\begin{align}
\theta_{\pm}=\frac{2}{R}\partial_{\pm}R,\label{expansion}
\end{align}
where 
\begin{align}
\partial_{+}& \ =\  \frac{\partial}{\partial\xi^{+}}=-\sqrt{2}\left[\frac{\partial_{t}}{N}-\frac{\partial_{r}}{a(t)}\right], \notag \\ 
\partial_{-}& \ =\ \frac{\partial}{\partial\xi^{-}}=-\sqrt{2}\left[\frac{\partial_{t}}{N}+\frac{\partial_{r}}{a(t)}\right],
\label{p+p_}
\end{align}
for these geodesics, which measure whether the bundle of null rays normal to
the sphere is diverging $(\theta_{\pm}>0)$ or converging $(\theta_{\pm}<0)$
\cite{MS}. Introducing the new parameter $\Theta(t,r):=\theta_{+}\theta_{-}$,
we get 
\begin{equation}
\Theta\ =\ \frac{8}{R^{2}}\left(\frac{\dot{R}^{2}}{N^2}-1\right)\ ,
\label{theta1-q}
\end{equation}
which in terms of the phase space variables, and working on the comoving gauge, by fixing $N=1$, can be written as
\begin{equation}
\Theta(c, {\sf p})\ =\ \frac{8}{r^{2}|{\sf p}|}\left(\frac{r^2}{\gamma^2}c^{2}-1\right)\ .
\label{theta1-q2}
\end{equation}
Using Eq.~(\ref{theta1-q}), the spacetime is said to be respectively,
\emph{trapped}, \emph{untrapped} or \emph{marginally trapped}, depending
on whether 
\begin{equation}
\Theta(t,r)>0,\ \ \ \ \ \Theta(t,r)<0,\ \ \ \ \ \ \Theta(t,r)=0\ .\label{sp}
\end{equation}
The third case in Eq. (\ref{sp}) characterises the outermost boundary
of the trapped region, namely the ``apparent horizon'', which corresponds
to the equation $\dot{R}^{2}=1$. Note that Eq.~(\ref{theta1-q})
is also to be thought of as a function in phase space, for every fixed
shell $r$. Specifically, we will have particular interest in the
boundary shell, $r=r_{b}$, which bounds the support of matter. For
that case, we can define 
\begin{equation}
\Theta_{b}(t):=\Theta(t,r_{b})=8\left(\frac{\dot{a}^{2}}{a^{2}}-\frac{1}{a^{2}r_{b}^{2}}\right).\label{THETAboundary}
\end{equation}
Since we are mainly interested in the eventual trapped surfaces formation,
due to the interior spacetime gravitational collapse, we assume that
the star is not trapped from the initial configuration at $t_{0}$;
in other words, $\Theta(t_{0},r)<0$ for all shells $0<r<r_{b}$.

At the classical level, it is possible to solve the Hamilton equation
$\dot{\phi}=\pi_{\phi}/|{\sf p}|^{3/2}$, analytically, which has
a general solution 
\begin{equation}
\phi\ =\ \pm\sqrt{\frac{3}{16\pi G}}\ln\frac{|{\sf p}|}{|{\sf p}_{0}|}+\phi_{0}\ .\label{classF3}
\end{equation}
In Eq. (\ref{classF3}), $(\phi_{0},{\sf p}_{0})$ are defined as
integration constants describing the initial conditions for the collapsing
star at the $t=t_{0}$ space slice. We will see in the next section
that, for $\phi\rightarrow\infty$, we have ${\sf p}=0$, i.e., the
volume of the interior region vanishes; however, since the matter
must be contained in such a region, the energy density of the cloud
diverges, producing a physical singularity. We will see that once
quantum corrections are taken into account, this scenario displays
a rather different physical outcome.

To model the exterior geometry, we choose a metric of the Vaidya family%
\footnote{This is a generalisation of the Schwarzschild metric, which accounts
for the possible matter emissions and the astrophysical realistic
case of a star surrounded by a radiating zone \cite{Vaidya,Vaidya2,Vaidya3}.%
}. Written in advanced Eddington-Finkelstein coordinates $(\mathrm{v},r_{\mathrm{v}},\theta,\phi)$,
it has the form \cite{Vaidya,Vaidya2,Vaidya3} 
\begin{align}
g_{\mu\nu}^{+}dx^{\mu}dx^{\nu} & \ =\   -\left(1-\frac{2GM(r_\mathrm{v},\mathrm{v})}{r_{\mathrm{v}}}\right)d\mathrm{v}^{2} \notag  \\ 
 & \  \  \   \  \  \  -2d\mathrm{v}dr_{\mathrm{v}} +r_{\mathrm{v}}^{2}d\Omega^{2},\label{EXTmetric}
\end{align}
where $M(r_\mathrm{v}, \mathrm{v})$ is a generic function of $r_\mathrm{v}$ and $\mathrm{v}$, which
is fixed by matching the Eq. (\ref{EXTmetric}) with Eq. (\ref{INTmetric})
at the boundary $r=r_{b}$ (for a discussion, see \cite{TachL,4}).
The matching conditions are defined by matching the area radius at
the boundary~$\Sigma$ \cite{2}: 
\begin{equation}
r_{\mathrm{v}}(\mathrm{v})\ \overset{\Sigma}{=}\ R(r_{b},t)=r_{b}a(t),\label{match1}
\end{equation}
together with the first and second fundamental forms 
\begin{eqnarray}
\left(\frac{d\mathrm{v}}{dt}\right)_{\Sigma} & = & \frac{R_{,r}+r_{b}\dot{a}}{1-\frac{F}{R}}\;,\label{match2}\\
F(t,r_{b}) & = & 2M(r_{\mathrm{v}},\mathrm{v})G\;,\label{match3}\\
GM(r_{\mathrm{v}},\mathrm{v})_{,r_{\mathrm{v}}} & = & \frac{F}{2R}+r_{b}^{2}a\ddot{a}\;.\label{match4}
\end{eqnarray}
It should be noted that the singularity formation at $a=0$ is independent
of these matching conditions.

Matching the exterior Vaidya geometry to the interior spacetime region
plays two important roles in a collapsing process: In the one hand,
it allows the matter to be radiated away as the collapse evolves;
on the other hand, it enables the study of horizons formation and
their evolution during the collapse. The second aspect is particularly
important; indeed, the formation of a black hole as the end state
of a collapsing star indicates that there exists a moment when an
apparent horizon develops inside the cloud, so that, whole the matter
collapses inside that horizon. Otherwise, if the final state is not
a black hole, the trapped surfaces never develop at any stage of the
collapse, and hence, no apparent horizons form inside the star. In
this paper we focus on the question of whether or not trapped surfaces
can form in the interior region, once quantum gravity corrections
are taken into account.

\section{Semiclassical scenario}
\label{LQG-Collapse}

On this section, we discuss the quantum gravity induced corrections
to the classical setting which was introduced in the previous section.
To do this, we implement an effective scenario provided by the holonomy
corrections imported from LQG (cf. see Ref. \cite{A-P-S2006} for details) for the spherically symmetric model (\ref{INTmetric}).

\begin{figure*}
\includegraphics[height=2.1in]{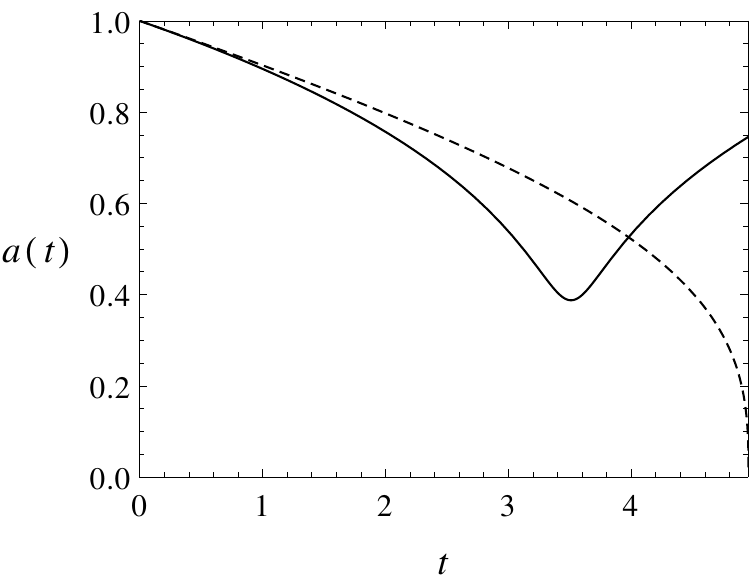} \qquad{} \includegraphics[height=2.12in]{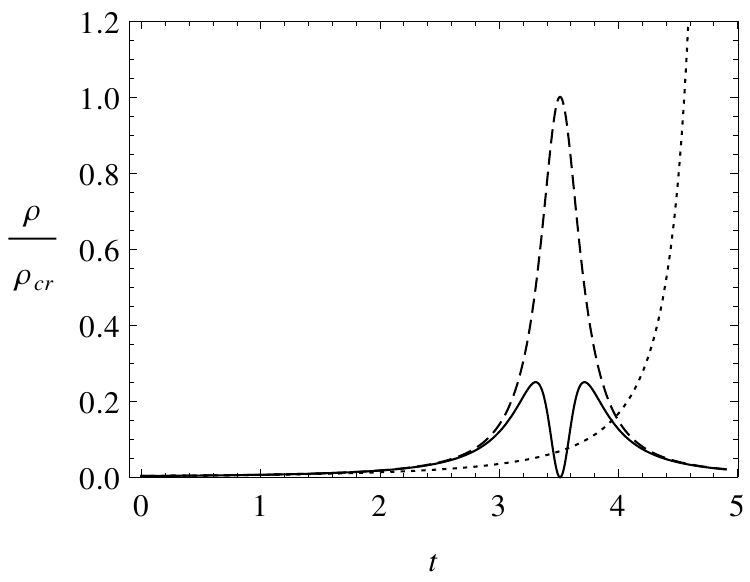}
\caption{%
{\footnotesize{{The left plot shows the time evolution of scale
factor $a(t)$ in the classical (dotted curve) and semiclassical (solid
curve) regimes. The right plot shows the classical (dotted curve)
and semiclassical (dashed curve) energy densities $\rho$ of scalar
field; the solid curve shows the behaviour of effective energy density
$\rho_{{\rm eff}}$ in the semiclassical regime. We have used the value of parameters
$G=c_{\mathrm{light}}=1$, and $\pi_{\phi}=10~000$.}}}%
}
\label{F-scalef} 
\end{figure*}

In LQC, the holonomization process must be implemented
to approach as much as possible the full theory of LQG. The hamiltonian constraint
can be derived through an embedding method \cite{LQC-AS}, where the $\mu_0$ quantization \cite{A-P-S2006} and $\mu$ improved quantization \cite{A-P-S} are well known examples. In the present work we will adopt $\mu$ quantization scheme. This choice is related to the fact that the LQC dynamics following the  $\mu_0$ quantization prescription, presented several unphysical features \cite{sLQC}.
In this scheme, the algebra generated by the holonomy of the phase space variables
$c$ is the algebra of the almost periodic function of $c$, i.e.,
$e^{i\mu c/2}$ (where $\mu$ is inferred as the kinematical length
of the square loop, since its dimension is similar to that of a length);
these functions together with ${\sf p}$, constitute the fundamental
canonical variables in the quantum theory \cite{A-B-L}. The procedure
consists in replacing $c^{2}$ by a $\sin^{2}(\mu c)/\mu^{2}$ in
Eq.~(\ref{Hamil-gr}); hence we have \cite{A-P-S2006,Taveras}
\begin{equation}
C_{{\rm eff}}\ =\ -\frac{3}{4\pi G\mu^{2}\gamma^{2}}\sqrt{|{\sf p}|}\sin^{2}(\mu c)+C_{{\rm matt}}\ .\label{EFFham}
\end{equation}
The dynamics of the fundamental variables is obtained by solving the
system of Hamilton equations; i.e. \cite{A-P-S2006}, 
\begin{align}
\dot{{\sf p}}  \ =\ \{{\sf p},C_{{\rm eff}}\} &\ =\ -\frac{8\pi G\gamma}{3}\frac{\partial C_{{\rm eff}}}{\partial c}\notag\\
 & \ =\ \frac{2\sqrt{|{\sf p}|}}{\gamma\mu}\sin(\mu c)\cos(\mu c).\label{HAMeqs-1}
\end{align}
If we consider the Hamilton equations (\ref{HAMeqs-1}) and the vanishing
Hamiltonian constraint (\ref{EFFham}), we can define a modified Friedmann
equation, $H=(\dot{a}/a)=(\dot{{\sf p}}/4{\sf p})$ \cite{A-P-S2006}:
\begin{equation}
\frac{\dot{a}^{2}}{a^{2}}\ =\ \frac{8\pi G}{3}\rho\left(1-\frac{\rho}{\rho_{\text{cr}}}\right)=:\frac{8\pi G}{3}\rho_{{\rm eff}}\ ,\label{Friedmann-eff}
\end{equation}
where $\rho_{\text{cr}}:=\sqrt{3}\rho_{\text{Pl}}/(16\pi^{2}\gamma^{3})\approx0.41\rho_{\text{Pl}}$
and $\rho_{\text{Pl}}$ is the Planck energy density. Eq. (\ref{Friedmann-eff})
implies that the classical energy density $\rho$ is limited by the
interval $\rho_{0}<\rho<\rho_{\text{cr}}$, which presents an upper
bound at $\rho_{\text{cr}}$. Notice that $\rho_{0}\ll\rho_{\text{cr}}$
is the energy density of the star at the initial configuration ($t=0$),
with $\rho_{0}=\pi_{\phi}^{2}/(2a_{0}^{6})$ and $a_{0}=a(0)$. Hence,
the quantum geometry effects are associated to an energy density modification,
proportional to $-\rho^{2}$, which becomes important when the energy
density becomes comparable to $\rho_{\text{cr}}$. Furthermore, in
the limit $\rho\rightarrow\rho_{\text{cr}}$, the Hubble rate vanishes;
the classical singularity is thus replaced by a bounce (cf. figure\footnote{In numerical studies in this work, we have used Mathematica (http://www.wolfram.com).}
\ref{F-scalef}). 
%\textcolor{blue}{In this sense, the quantum geometry effects resolve
%the classical singularity developing at the late time evolution of
%the collapse\footnote{\textcolor{red}{Again, the authors must be careful about the affirmations related to the singularity resolution. 
%These results indicate that it is in fact the expected situation, but they are not a rigorous proof of singularity resolution.}}}. 
Notice that, in the limit $\rho\ll\rho_{\text{cr}}$,
the standard Friedmann equation is recovered.

\begin{figure}
\centering\includegraphics[height=2.2in]{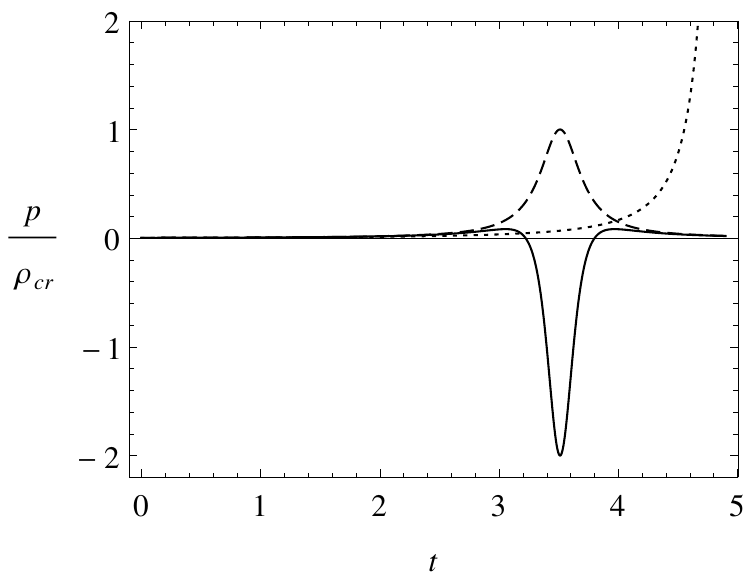} \caption{%
{\small{{This plot represents the behaviour of the pressure in the
classical and semiclassical regimes for the values of parameters $G=c_{\mathrm{light}}=1$,
and $\pi_{\phi}=10~000$. In the classical collapse, the matter pressure
(dotted curve) increases and diverges towards the singularity. In
the semiclassical regime, the matter pressure (dashed curve) increases
and reaches to a maximum $p=\rho_{{\rm cr}}$ at the bounce, wheras
the effective pressure $p_{\textrm{eff}}$ (solid curve) decreases
and takes a minimum super negative value $p_{\textrm{eff}}(\rho_{\textrm{cr}})=-2\rho_{\textrm{cr}}$
at the bounce.}}}%
}

\label{p-eff-1} 
\end{figure}

From the Raychaudhuri equation we can define the effective pressure
for the massless scalar field as \cite{LQC-AS,A-P-S2006} 
\begin{equation}
p_{\mathrm{eff}}\ :=\ \rho\left(1-3\frac{\rho}{\rho_{\mathrm{cr}}}\right).\label{press-eff}
\end{equation}
Figure~\ref{p-eff-1} represents the behaviour of the pressures $p$
and $p_{{\rm eff}}$ in Eq.~(\ref{press-eff}) conveniently scaled
with the critical density $\rho_{\textrm{cr}}$. In the semiclassical
regime, the matter pressure, $p=\rho=\pi_{\phi}^{2}/2a^{6}$, increases
during the collapse (see dashed curve in figure~\ref{p-eff-1}),
but remains finite until the bounce where it reaches a maximum at
$p_{{\rm cr}}=\rho_{{\rm cr}}$. The effective pressure (solid curve)
is positive initially, then as energy density increases, $p_{{\rm eff}}$
decreases until it vanishes at $\rho=\rho_{{\rm cr}}/3$. In the range
$\rho_{{\rm cr}}/3<\rho<\rho$, the effective pressure evolves negatively
until the bounce where it takes the super negative value $p_{\textrm{eff}}(\rho_{\mathrm{cr}})=-2\rho_{\textrm{cr}}$
at the bounce. This indicates that, in the herein homogeneous and
isotropic collapsing model, the singularity resolution is associated
with the violation of (effective) energy conditions (e.g., $\rho_{{\rm eff}}+p_{{\rm eff}}<0$),
which suggests that the quantum gravity effects provide a repulsive
force at the very short distances \cite{NegativePress}. This feature
may also result in a strong burst of outward energy flux in the semiclassical
regime.

\begin{figure}
\centering\includegraphics[height=2.2in]{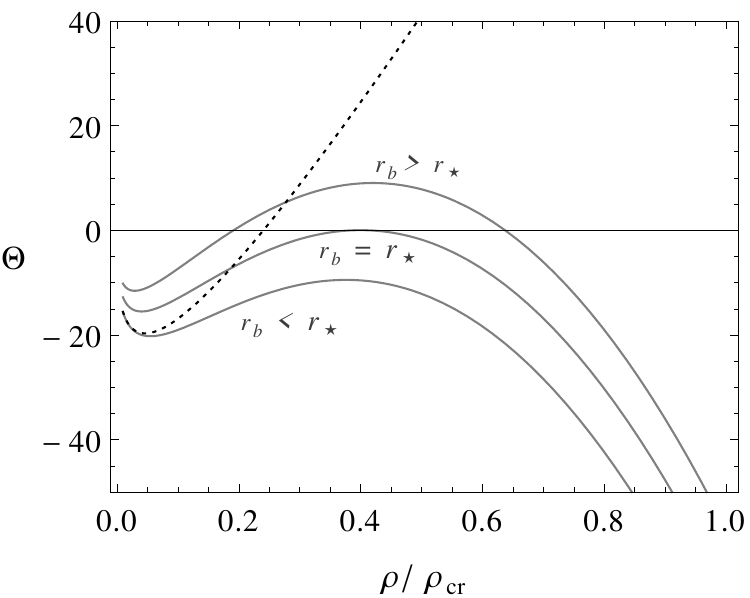} \caption{%
 {\footnotesize{{Behaviours of $\Theta_{b}(\rho)$ in the classical
(dotted curve), and semiclassical (solid curves) regimes for different
values of $r_{b}$. We have used the value of parameters
$G=c_{\mathrm{light}}=1$, and $\pi_{\phi}=10~000$.}}}%
}

\label{F-thetaa} 
\end{figure}

To discuss the trapped surfaces dynamics, particular importance is
played by the function $\Theta_{b}$ defined in Eq.~(\ref{THETAboundary}).
Therein, by replacing $\dot{a}/a$ with the effective Friedmann equation
(\ref{Friedmann-eff}) we have 
\begin{equation}
\Theta_{b}\ =\ \frac{64\pi G}{3}\rho\left(1-\frac{\rho}{\rho_{\text{cr}}}\right)-\frac{8}{a^{2}r_{b}^{2}}\ .\label{theta2-q2}
\end{equation}
We will assume that the cloud is initially untrapped, and thus for
$\rho\ll\rho_{{\rm cr}}$, we have that $\Theta_{b}(t=0)$ is negative.
Now, we can study the behaviour of the effective $\Theta_{b}$ as
a function of the energy density $\rho$. Let us rewrite Eq.~(\ref{theta2-q2})
by setting $X:=\rho/\rho_{{\rm cr}}$ as 
\begin{equation}
\Theta_{b}(X)\ =\ AX\left(1-X\right)-BX^{1/3}\ ,\label{theta2-q2-X}
\end{equation}
where $A:=(64\pi G/3)\rho_{{\rm cr}}$ and $B:=8(2\rho_{{\rm cr}}/\pi_{\phi}^{2})^{1/3}/r_{b}^{2}$
are constants. The behaviours of $\Theta_{b}$, with respect to $X$,
for the different choices of the initial conditions, are sketched
in figure \ref{F-thetaa}. Therein, the solid curves represent the
trajectories provided by the semiclassical gravitational collapse;
whereas the dotted curve shows the classical trajectories (which coincides
with the semiclassical ones for $X\ll1$). An equation defining the
apparent horizon for the effective geometry can be obtained by equating
(\ref{theta2-q2-X}) to zero. So, we obtain 
\begin{equation}
X^{2}(1-X)^{3}-\left(\frac{B}{A}\right)^{3}=0\ .\label{AH-EQ}
\end{equation}
To solve this last equation, we compute the values
of energy density at which the apparent horizons form. This corresponds
to the intersections of the $\Theta_{b}$ curve with the horizontal
axe in figure \ref{F-thetaa}. Therefore, depending on the initial
conditions, in particular on the choice of the boundary radius $r_{b}$,
three cases can be evaluated, which correspond to \emph{no} apparent
horizon formation, one and two horizons formation. Notice that, denoted
by a dotted curve, only one horizon can form classically. Let us to
be more precise as follows.

In the one hand, the modified Friedmann equation (\ref{Friedmann-eff})
allows to determine the energy density at which speed of the collapse,
$|\dot{a}|$, reaches its maximum. From Eq.~(\ref{Friedmann-eff})
we can present $|\dot{a}|$ as 
\begin{equation}
|\dot{a}|\ =\ \sqrt{A_{0}}X^{1/3}(1-X)^{1/2}\ ,
\end{equation}
where $A_{0}:=(8\pi G/3)(\pi_{\phi}^{2}\rho_{{\rm cr}}^{2}/2)^{\frac{1}{3}}$
is a constant. It follows that for the energy density $\rho=(2/5)\rho_{{\rm cr}}$,
the speed of the collapse is maximum at 
\begin{equation}
|\dot{a}|_{\text{max}}\ =\ \sqrt{\frac{3}{5}A_{0}}\left(\frac{2}{5}\right)^{\frac{1}{3}}.\label{maxadot}
\end{equation}
The scale factor $a_{\mathrm{max}}$, corresponding to $|\dot{a}|_{\text{max}}$
reads $a_{\mathrm{max}}=(5\pi_{\phi}^{2}/4\rho_{{\rm {cr}}})^{1/6}$.
Notice that this value is independent of $r_{b}$, therefore, it is
the same for any shell. The minimum value of the scale factor, $a_{\text{cr}}$,
is fixed by the requirement that the Hubble rate vanishes, i.e., $\rho=\rho_{\text{cr}}$,
when the collapse hits a bounce; at this point we have that $a_{\text{cr}}=(\pi_{\phi}^{2}/2\rho_{{\rm {cr}}})^{1/6}=(2/5)^{1/6}a_{\text{max}}$.

On the other hand, by setting $\Theta_{b}=0$ in Eq. (\ref{theta1-q})
we get $\dot{R}^{2}=1$, so that, we can determine the speed of the
collapse, $|\dot{a}|_{\mathrm{AH}}=1/r$, for any shell $r$, at which
an horizon can form; in particular, for the boundary shell, this gives
$|\dot{a}|_{\mathrm{AH}}=1/r_{b}$. When the speed of the collapse,
$|\dot{a}|$, reaches the value $1/r_{b}$, then an apparent horizon
forms. Thus, if the maximum speed $|\dot{a}|_{\text{max}}$ is lower
than the critical speed $|\dot{a}|_{\mathrm{AH}}$, no horizon can
form. Let us introduce a radius $r_{\star}$, as 
\begin{equation}
r_{\star}\ :=\ \frac{1}{|\dot{a}|_{\text{max}}}\ .
\end{equation}
We see that $r_{\star}$ determines a \emph{threshold radius} for
the horizon formation in the scalar field collapse with the momentum
$\pi_{\phi}$; if $r_{b}<r_{\star}$, then \emph{no} horizon can form
at any stage of the collapse. The case $r_{b}=r_{\star}$ corresponds
to the formation of a dynamical horizon at the boundary of the two
spacetime regions \cite{dynHorizon,dynHorizon2}. Finally, for the case $r_{b}>r_{\star}$
two horizons will form, one inside and the other outside the collapsing
matter.

\section{Semiclassical outcomes of the collapse}

\label{Collapse-outcome}

So far, we have analysed the interior collapsing
spacetime in the presence of the quantum gravity effects. This quantum
effects are expected to be carried out to the exterior geometry through
the matching conditions applied on the boundary $r_{b}$ of two regions.
In the following, we will focus on the main physical consequences
that can emerge from this scenario in order to predict the possible
exterior geometry for the collapse.

\begin{figure*}
\includegraphics[height=2.2in]{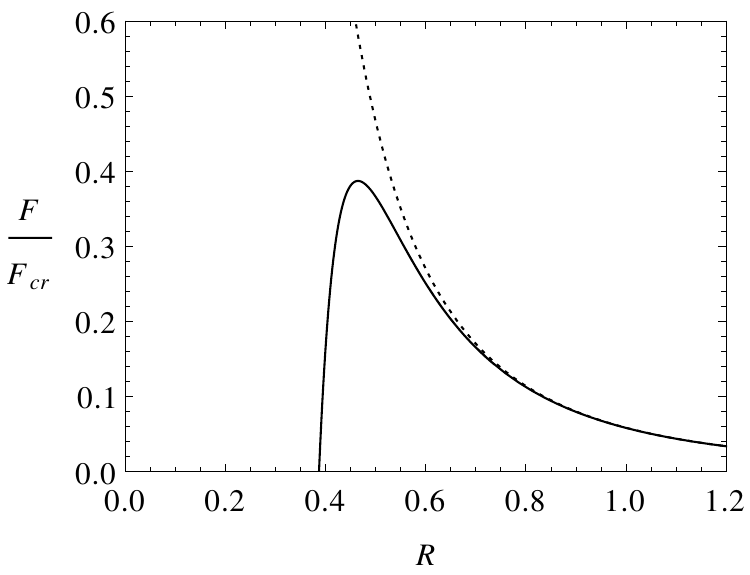} \qquad{} \includegraphics[height=2.2in]{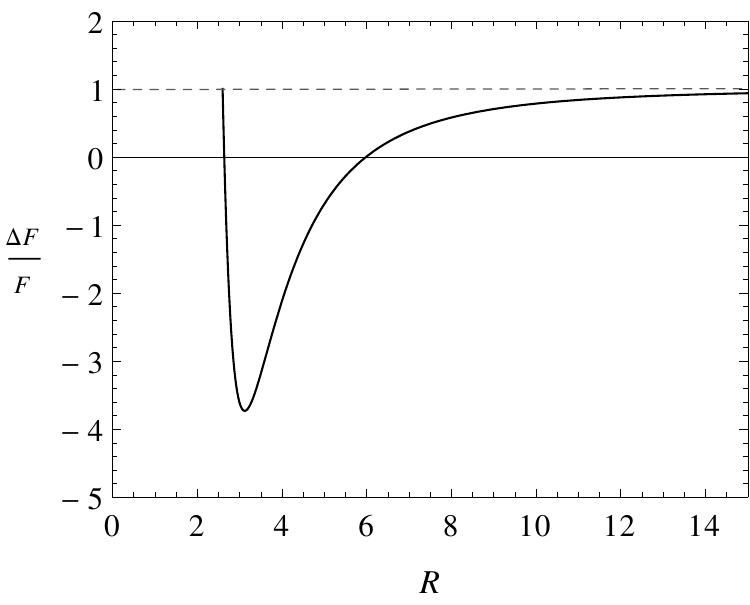}
\caption{%
 {\small{{The left plot represents the behaviour of classical (dotted
curve) and effective (solid curve) mass function during the collapse.
The right plot shows the behaviour of the mass loss $\Delta F/F$,
as a function of area radius $R$. We have used the value of parameters
$G=c_{\mathrm{light}}=1$, and $\pi_{\phi}=10~000$. }}}
}

\label{F-lum} 
\end{figure*}

The classical Friedmann equation corresponds to the last relation
in the classical Einstein's field equation (\ref{einstein-1}), which
can be written in terms of the mass function as $H^{2}=F/R^{3}$.
Consequently, and since in the semiclassical regime the Friedmann
equation is modified to Eq.~(\ref{Friedmann-eff}), this might imply
a modification of the mass function defined by Eq. (\ref{einstein-1}).
In other words, we can introduce an effective mass function $F_{\text{eff}}$
corresponding to the modified Friedmann equation (\ref{Friedmann-eff})
as 
\begin{equation}
F_{\text{eff}}\ =\ \frac{8\pi G}{3}\rho_{\text{eff}}R^{3}\  =\  \frac{8\pi G}{3}\rho R^{3}\left(1-\frac{\rho}{\rho_{\rm cr}}\right).\label{mass-eff}
\end{equation}
This describes an effective geometry on which the phase space trajectories
are considered to be classical, whereas the matter content is assumed
to be modified by quantum gravity effects. In the classical limit,
as $\rho_{\text{eff}}\rightarrow\rho$, the effective mass function
reduces to the classical one given by Eq.~(\ref{mass-class}). In the interior
semiclassical region, since $\rho_{0}<\rho<\rho_{\text{cr}}$, so
both $F$ and $F_{{\rm eff}}$ remain finite during the collapse.
Using the relations $F=(8\pi G/3)\rho R^{3}$ and $\rho/\rho_{\textrm{cr}}=F^{2}/F_{\textrm{cr}}^{2}$  (for a massless scalar field),
it is convenient to rewrite Eq. (\ref{mass-eff}) as % \textcolor{red}{[explain more, how we get Eq. (27) and (28)?]}
\begin{equation}
F_{\text{eff}}\ =\ F\left(1-\frac{F^{2}}{F_{\mathrm{cr}}^{2}}\right),\label{mass-effb}
\end{equation}
in which we have defined  %, %for a massless scalar field we have used 
%\begin{equation}
%\frac{\rho}{\rho_{\textrm{cr}}}\ =\ \frac{F^{2}}{F_{\textrm{cr}}^{2}}\ ,
%\end{equation}
%with 
$F_{\mathrm{cr}}:=8\pi G\pi_{\phi}^{2}r_{b}^{3}\sqrt{\rho_{\textrm{cr}}}/3\sqrt{2}$. Notice that $F_{\rm cr}$ is a function of the phase space variable $\pi_\phi$; 
since,  for a massless scalar field, $\pi_\phi$ is a constant of motion, fixed by the initial conditions,  $F_{\mathrm{cr}}$ becomes a constant for any  shell (with a specific choice of  $r_b$) and is determined at the initial configuration of the collapse.
Eq. (\ref{mass-effb}) shows that, the mass function $F$ is allowed
to evolve in the interval $F_{0}<F<F_{\mathrm{cr}}$ along with the
collapse dynamical evolution. Consequently, the effective mass function
$F_{{\rm eff}}$ increases from the initial value
$\sim F_{0}$ (for $\rho\ll\rho_{\textrm{cr}}$) and reaches a maximum
at $F_{{\rm cr}}/\sqrt{3}$; then, it starts decreasing and vanishes
at $F_{{\rm cr}}$ (cf. see left plot in figure~\ref{F-lum}). In
addition, it should be noticed that, classically trapped surfaces
form when $F>R$ at some points during the collapse and $F$ diverges
at the singularity. Nevertheless, in the presence of quantum effects,
this situation is different. For the choice of $r_{b}<r_{\star}$,
the effective mass function remains $F_{{\rm eff}}<R$, so that, no
trapped surface forms; if $r_{b}\geq r_{\star}$, then $F_{{\rm eff}}\geq R$
and trapped surfaces form during the collapse.

For a collapsing star whose initial boundary radius $r_{b}$ is less
than $r_{\star}$, we study the resulting mass loss due to the semiclassical
modified interior geometry. Let us designate the initial mass function
at scales $\rho\ll\rho_{{\rm cr}}$, i.e, in the classical regime,
as $F_{0}=(8\pi G/3)\rho_{0}R_{0}^{3}$, where $\rho_{0}=\pi_{\phi}^{2}/2a_{0}^{6}$,
and for $\rho\lesssim\rho_{{\rm cr}}$ (in the semiclassical regime)
we have $F_{\text{eff}}$ given by Eq. (\ref{mass-effb}). Then, the
(quantum geometrical) mass loss, $\Delta F/F_{0}$ (where $\Delta F=F_{0}-F_{\text{eff}}$),
for any shell is provided by the following expression: 
\begin{align}
\frac{\triangle F}{F(a_{0})}\   =\ 1-\frac{F_{\text{eff}}}{F_{0}}\ =\ 1-\sqrt{\frac{\rho}{\rho_{0}}}\left(1-\frac{\rho}{\rho_{{\rm cr}}}\right)\ .\label{mass-loss-eff}
\end{align}
As $\rho$ increases the mass loss decreases positively until it vanishes
at a point. Then, $\Delta F/F$ continues decreasing (negatively)
until it reaches to a minimum at $\rho=\rho_{{\rm cr}}/3$. Henceforth,
in the energy interval $\rho_{{\rm cr}}/3<\rho<\rho_{{\rm cr}}$,
the mass loss increases until the bouncing point at
$\rho\rightarrow\rho_{\text{cr}}$, where $\Delta F/F\rightarrow1$;
this means that the quantum gravity corrections, applied
to the interior region, give rise to an outward flux of energy near
the bounce in the semiclassical regime.

It is worthy to mention that, when an inverse triad correction is
applied to the collapsing system (with a scalar field \cite{4}, or
a tachyon field \cite{TachL}, as matter sources), the (quantum) modified
energy density decreases as collapse evolves. Whence, as the collapsing
cloud approaches the center (with a vanishing scale factor, where
the classical singularity is located) the energy density reaches its
minimum value, whereas the mass loss tends to one. In the holonomy
corrected semiclassical collapse herein, the energy density increases
and reaches to a maximum value $\rho_{\text{cr}}$ at the bounce (with
a finite non-zero volume). Nevertheless, the dynamics of the collapse
is governed by an effective energy density which decreases close to
the bounce and vanishes at $a=a_{\textrm{cr}}$. Consequently,
the effective mass function also decreases and vanishes at the bounce,
which happens at $t_{\textrm{cr}}<t_{{\rm sing}}$ (with $t_{{\rm sing}}$
being the time when the classical singularity is reached).

If the initial condition for the collapsing star is such that $r_{b}\geq r_{\star}$,
then a black hole will form at the collapse final state. We will now
analyse a possible prediction for the exterior geometry of the collapsing
system in this case. The total mass measured by an asymptotic observer
is given by $m_{\mathrm{ext}}=m_{M}+m_{\phi}$, where $m_{M}$ is
the total mass in the generalized Vaidya region, and $m_{\phi}=\int\rho dV$
is the interior mass related to the scalar field $\phi$. Since the
matter related to $m_{M}$ is not specified in the
exterior Vaidya geometry in our model, we just focus on a qualitative
analysis of behaviour of the horizon close to the matter shells.

From the matching conditions (\ref{match1})-(\ref{match4}), we can
get the information regarding the behaviour of trapping horizons in
the exterior region. Indeed, when the relation $2M(\mathrm{v},r_{\mathrm{v}})G=r_{\mathrm{v}}$
is satisfied at the boundary, trapped surfaces will form in the exterior
region close to the matter shells. On classical geometry,
the boundary function, ${\cal F}=(1-2M(\mathrm{v})G/r_{\mathrm{v}})$,
becomes negative for the trapped region and vanishes at the apparent
horizon. Therefore, the equation for event horizon is given at the
boundary of the collapsing body by ${\cal F}|_{\Sigma}=0$. Nevertheless,
in the semiclassical regime, the boundary function is expected to
be modified by employing the matching conditions due to the fact that
the interior spacetime was modified by the quantum gravity effects.
Using the conditions (\ref{match1}) and (\ref{match3}), we have
that $2M(\mathrm{v},r_{\mathrm{v}})G/r_{\mathrm{v}}=F(t)/R(t)$ at
the boundary surface $\Sigma$ with $r=r_{b}$. Since the mass function
is modified as in the Eq. (\ref{mass-eff}) in the semiclassical regime,
therefore, the mass $M(\mathrm{v})$ is also modified as $\tilde{M}(\mathrm{v})=F_{{\rm eff}}/2G$
at $\Sigma$: 
\begin{equation}
\tilde{M}(\mathrm{v})\ =\ M-\frac{M^{3}}{M_{\mathrm{cr}}^{2}}\ ,\label{bounF-eff}
\end{equation}
where $2GM_{{\rm cr}}:=F_{{\rm cr}}=const$. Eq. (\ref{bounF-eff})
shows that the quantum gravity induced effects leads to a modification
of the boundary function by a cubic term $M^{3}$. By substituting
the classical mass function with $F=(8\pi G/3)\rho R^{3}$, we can
rewrite the Eq. (\ref{bounF-eff}) as 
\begin{equation}
\tilde{M}(R)\ =\ \frac{C}{R^{3}}-\frac{D}{R^{9}}\ ,\label{bounF-eff2}
\end{equation}
where $C:=(2\pi/3)\pi_{\phi}^{2}r_{b}^{6}$, and $D:=C\pi_{\phi}^{2}r_{b}^{6}/(2\rho_{\mathrm{cr}})$
are constants. Eq. (\ref{bounF-eff2}) represents a non singular,
exotic black hole geometry. Notice that, the effective exterior function
${\cal F}_{{\rm eff}}=(1-2G\tilde{M}/R)$ in the classical limit,
where $\rho_{\mathrm{cr}}\rightarrow\infty$, tends to ${\cal F}=1-2GC/R^{4}$,
which represents a classical singular black hole geometry \cite{VHussain}.
In addition, as we expected, in the presence of a nonzero matter pressure
(of the massless scalar field) at the boundary, the (homogeneous)
interior spacetime is not matched with an empty (inhomogeneous) Schwarzschild
exterior \cite{homo-Boj}.

\section{Conclusions and discussion}

\label{discussion}

We have considered a spherically symmetric and homogeneous spacetime
for a gravitational collapse whose matter content is a massless scalar
field. The homogeneous interior is matched to an exterior Vaidya geometry.
We employed loop quantum gravity to investigate the quantum gravity
effects on the fate of the collapse. We subsequently studied the interior
spacetime within the effective theory of LQC. There are two types
of corrections that are considered in effective studies in LQC; \emph{holonomy}
and \emph{inverse triad} corrections. In this paper, we focused on
holonomy corrections applied to the interior region of the collapse.
It was shown that loop quantum effects remove the classical singularity
arising at the end state of the collapse, and replace it by a bounce.
Furthermore, we investigated the evolution of the trapped surfaces
emerging from the semiclassical interior spacetime. The physical modifications
related to the semiclassical regime provided three cases for the trapped
surfaces formation, depending on the initial conditions of the collapsing
star. In particular, our solutions showed that, if the initial boundary
radius of the collapsing cloud is less than a threshold radius, namely
$r_{\star}$, \emph{no} horizon forms during the collapse, whereas
for the radius equal and larger than the $r_{\star}$, one and two
horizons form, respectively. It is worthy to mention that, this scenario
is qualitatively similar to the model previously predicted from an
inverse triad correction \cite{homo-Boj}.

The interior semiclassical collapse can affect the exterior geometry
by imposing appropriate matching conditions on the boundary of two
regions. Therefore, an effective geometry emerged for the exterior
metric which allows the description of the physical consequences present
at the late time evolution of the collapse. For the case in which
no horizon forms, we have showed that, as the collapse evolves, the
energy density increases towards a maximum value $\rho_{\textrm{cr}}$
at the bounce. The energy density growth of the matter cloud seems
to be accompanied by a negative mass loss, however, in the herein
semiclassical collapse the effective energy density decreases which
leads to an apparent positive mass loss near the bounce. This results
in a positive luminosity near the bounce and gives rise to an outward
energy flux from the interior region which may reach to the distant
observer. A similar scenario was considered in study of gravitational
collapse of a standard scalar field \cite{4} and a tachyonic field
\cite{TachL}, where, instead of the holonomy correction, an \emph{inverse
triad} modification was employed. The mass loss obtained therein,
was characterised by a reduction of the energy density and mass function
towards the centre of the star, which leaded to an outward energy
flux from the interior region and reaching the distant observer. In
addition, in the cases in which one or two horizons form, the resulting
exterior geometry corresponds to an exotic nonsingular black hole
which is different from the Schwarzschild spacetime \cite{homo-Boj,VHussain}.

The qualitative picture that emerges from our toy model was influenced
by the choice of an homogeneous interior spacetime. Nevertheless,
in a realistic collapsing scenario one has to employ a more general
inhomogeneous setting (see Ref.~\cite{Bojowald:2006,Campiglia:2007}
for recent development of techniques to handle inhomogeneous systems,
which gives promising indications on how to extend the simpler homogeneous
case). Furthermore, the effective theory that predicts a modified
homogeneous dynamics, for the interior spacetime, may also modify
the spacetime inhomogeneous structure \cite{Bojowald:2012}. In addition,
when we apply homogeneous techniques, it is only the interior spacetime
that carries quantum effects; whereas the outside spacetime is described
by a generalised Vaidya metric in the context of the general relativity.
Some quantum effects are transported to the outside, by imposing suitable
matching conditions at the boundary surface, which enter the Vaidya
solution effectively through a nonstandard energy-momentum tensor.
This procedure is also limited by the fact that the consideration
of a complete inhomogeneous quantization, for the exterior region,
may provide significant modifications to the spacetime structure.
These effects may not be captured for a general Vaidya mass in a spacetime
line element \cite{Bojowald:2011}. It is an interesting problem to
estimate which of the physical predictions based on the homogeneous
models could still be present when a complete inhomogeneous is considered.
Nonetheless, it is difficult to calculate such effects in these more
realistic models, due to the fact that we are still far from a complete
picture.

\section*{Acknowledgement}

The authors would like to thank M.~Bojowald, R.~Goswami, P.~Singh
and J.~Velhinho for the useful discussion and suggestions. They 
thank P. Vargas Moniz for careful reading the manuscript and also thank the unknown Referee for making
useful comments on our work. YT was supported by the Portuguese Agency Funda\c{c}\~ao
para a Ci\^encia e Tecnologia through the fellowship SFRH/BD/43709/2008. 
This research work was also supported by the grants PEst-OE/MAT/UI0212/2014 and CERN/FP/123609/2011.

%%-------------------------------------------------------------------------------------------------


\begin{thebibliography}{10}

\bibitem{Hawking} S. W. Hawking, G. F. R. Ellis, \textit{The Large
Scale Structure of Space-Time}  (Cambridge University Press, 1974).

\bibitem{RPenrose} R. Penrose, Phys. Rev. Lett. \textbf{14}, 57 (1965).

\bibitem{RPenrose2} S. W. Hawking Proc. Roy. Soc. Lond. A \textbf{300}, 187 (1967).
\bibitem{RPenrose2b} S. W. Hawking and R. Penrose Proc. Roy. Soc. Lond. A \textbf{314}, 529 (1970).

\bibitem{2} P. Joshi, \textit{Gravitational Collapse and Spacetime
Singularities} (Cambridge University Press, 2007).

\bibitem{LQG} A. Ashtekar, J. Lewandowski, \emph{Background Independent
Quantum Gravity: A Status Report}, Class. Quant. Grav. \textbf{21}:
R 53 (2004), {[}arXiv: gr-qc/0404018{]}.

\bibitem{LQG2} C. Rovelli, \emph{Quantum Gravity} (Cambridge University
Press, Cambridge, England, 2004). 
 
\bibitem{LQG3} T. Thiemann, \emph{Introduction to Modern Canonical
Quantum General Relativity}, (Cambridge University Press, Cambridge,
England, 2007).

\bibitem{Bojowald-Book} M. Bojowald, \emph{Canonical Gravity and
Applications: Cosmology, Black Holes, and Quantum Gravity} (Cambridge
University Press, 2010).

\bibitem{Bojowald:2002} M. Bojowald, Class. Quantum Grav. {\bf 19}, 2717 (2002). 

\bibitem{A-B-L} A. Ashtekar, M. Bojowald, J. Lewandowski, \emph{Mathematical
structure of loop quantum cosmology}, Adv. Theor. Math. Phys. \textbf{7},
233 (2003).


\bibitem{Thiemann2006} J. Brunnemann and T.  Thiemann,  Class. Quantum Grav.  {\bf 23}, 1395 (2006).

\bibitem{Thiemann2006b} J. Brunnemann and T.  Thiemann,  Class. Quantum  Grav.  {\bf 23}, 1429 (2006).


\bibitem{4} R. Goswami, P. S. Joshi and P. Singh, Phys. Rev. Lett.
\textbf{96}, 031302 (2006).

\bibitem{homo-Boj} M. Bojowald, R. Goswami, R. Maartens, P. Singh,
Phys. Rev. Lett. \textbf{95}, 091302 (2005). 

\bibitem{TachL} Y. Tavakoli, J. Marto, A. H. Ziaie, and P. Vargas
Moniz, Phys. Rev. D \textbf{87}, 024042 (2013).

\bibitem{RTibrewala} M. Bojowald, T. Harada, R. Tibrewala,
Phys. Rev. D \textbf{78}, 064057 (2008).

\bibitem{RTibrewala-b} M. Bojowald, J. D. Reyes, R. Tibrewala, Phys.
Rev. D \textbf{80}, 084002 (2009).


\bibitem{Modesto-1} L. Modesto, Phys. Rev. D \textbf{70}, 124009
(2004); {[}arXiv:gr-qc/ 0504043{]}.

\bibitem{Modesto-1b} L. Modesto,  Int. J. Theor. Phys. \textbf{47},
357 (2008).


\bibitem{VHussain} B. K. Tippett and V. Husain, Phys. Rev.
D \textbf{84}, 104031 (2011).

\bibitem{TachClass} Y. Tavakoli, J. Marto, A. Ziaie, and P. Vargas
Moniz, Gen. Rel. Grav. \textbf{45}, 819 (2013).

\bibitem{A-P-S2006} A. Ashtekar, T. Pawlowski, and P. Singh, Phys.
Rev. D \textbf{73}, 124038 (2006).

\bibitem{A-P-S} A. Ashtekar, T. Pawlowski, and P. Singh, Phys. Rev.
D \textbf{74}, 084003 (2006).

\bibitem{LQC-AS} A. Ashtekar and P. Singh, \emph{Loop quantum cosmology:
a status report}, Class. Quantum Grav. \textbf{28}, 213001 (2011).

\bibitem{inhomogeneous1}  M. Bojowald, Class. Quantum Grav. {\bf 21}, 3733 (2004). 
\bibitem{inhomogeneous2} V. Hussain and  O. Winkler, Class. Quantum Grav. {\bf 22}: L127-L134 (2005).
\bibitem{inhomogeneous3} M. Bojowald, Phys. Rev. Lett.  {\bf 95}: 061301 (2005).


\bibitem{Bojowald:2011} M. Bojowald, G. M. Paily, J. D. Reyes and
R. Tibrewala, Class. Quantum Grav. \textbf{28}, 185006 (2011).

\bibitem{LL1} L. D. Landau and E. M  Lifshitz, {\em The classical theory of fields}, p. 304 (1975).

\bibitem{SS-1}  S. Bhattacharya, R. Goswami and P. S. Joshi, Int. J. Mod. Phys. D {\bf 20}, 1123 (2011).
\bibitem{SS-1b}  S. Bhattacharya, Proceedings of JGRG19, Japan  2009; arXiv:1107.4112 {[}gr-qc{]}.
\bibitem{SS-1c}  S. Bhattacharya, R. Goswami, and P. S. Joshi. arXiv:0807.1985  [gr-qc].
 
\bibitem{Giambo} R.  Giamb\'o,  Class.  Quantum Grav. {\bf 22}, 2295 (2005).
 
\bibitem{sLQC} A. Ashtekar, A. Corichi and  P. Singh,  Phys. Rev. D {\bf 77} 024046 (2008).

\bibitem{MS} S. A Hayward, Phys. Rev. D \textbf{53}, 1938 (1996).

\bibitem{Vaidya} P. C. Vaidya,  Curr. Sci. {\bf 12}, 183 (1943). 
\bibitem{Vaidya2} P. C. Vaidya, Proc. Indian Acad. Sci. A. {\bf 33}, 264 (1951). 
\bibitem{Vaidya3} P. C. Vaidya,  Nature, {\bf 171}, 260 (1953).

\bibitem{Taveras} V. Taveras, IGPG preprint (2006).

\bibitem{dynHorizon} S. Hayward, Phys. Rev. D \textbf{49}, 6467 (1994).
\bibitem{dynHorizon2} A. Ashtekar and B. Krishnan, Phys. Rev. Lett. \textbf{89}, 261101 (2002).

\bibitem{NegativePress} P. Singh, Class. Quant. Grav. \textbf{22}, 4203 (2005).

\bibitem{Bojowald:2006} M. Bojowald, R. Swiderski, Class. Quantum Grav. \textbf{23}, 2129 (2006).

\bibitem{Campiglia:2007} M. Campiglia, R. Gambini and J. Pullin,
Class. Quantum Grav. \textbf{24}, 3649 (2007).

\bibitem{Bojowald:2012} M. Bojowald, 
%\emph{Quantum cosmology: effective theory}, 
Class. Quantum Grav. \textbf{29}, 213001 (2012).



\end{thebibliography}
\end{document}